# Amplified Spontaneous Emission and Random Lasing in MAPbBr₃ Halide Perovskite Single Crystals

Amplified Spontaneous Emission and Random Lasing in MAPbBr$_3$ Halide Perovskite Single Crystals

*Aleksei O. Murzin[1,*], Boris V. Stroganov[1], Carsten Günnemann[2], Samia Ben Hammouda[2], Anna V. Shurukhina[1], Maxim S. Lozhkin[1], Alexei V. Emeline[1], Yury V. Kapitonov[1,**]*

1 Saint Petersburg State University, 7/9 Universitetskaya Emb., 199034, St. Petersburg, Russia
2 Institut für Technische Chemie, Leibniz Universität Hannover, Callinstraße 3, 30167 Hannover, Germany
*E-mail: alekseymurzin10@gmail.com
** E-mail: yury.kapitonov@spbu.ru

Keywords: halide perovskites, amplified spontaneous emission, random lasing

Halide perovskites are a promising optical gain media with high tunability and simple solution synthesis. In this study, two gain regimes, namely amplified spontaneous emission and random lasing, are demonstrated in same MAPbBr$_3$ halide perovskite single crystal. For this, photoluminescence is measured at a temperature of 4 K with pulsed femtosecond pumping by UV light with a 80 MHz repetition rate. Random lasing is observed in areas of the sample where a random resonator was formed due to cracks and crystal imperfections. In more homogeneous regions of the sample, the dominant regime is amplified spontaneous emission. These two regimes are reliably distinguished by the line width, the mode structure, the growth of the intensity after the threshold, and the degree of polarization of the radiation. The spectral localization of the stimulated emission well below the bound exciton resonance raises a question concerning the origin of the emission in halide perovskite lasers.

## 1. Introduction

Halide perovskites have emerged recently as a new class of photovoltaic and optoelectronic semiconductor materials. In short time the effectiveness of halide perovskite solar cells has reached 25.2% [1]. In addition to the excellent light-absorption properties, these materials also exhibit good light-emitting properties and could be used as a gain media for lasers. At the moment, most of the works are devoted to the study of amplified spontaneous emission (ASE) and lasing with pulsed optical pumping. The resonators used for lasing can be divided into three groups: random resonators, supporting random lasing (RL) [2-6]; optical resonators based on the refractive index contrast between the synthesized perovskite material and air, such as micro- and nano- cubes [7-9], rods [10,11], wires [12-17], platelets [18-20], micro pyramids [21] and microspheres [22-24]; and external resonators based on distributed Bragg reflectors [25-28], capillary-like microcavities [29], photonic crystals [30, 31], gratings for distributed feedback lasers [32-38], and nanolithographically fabricated micro disks [39-41]. The laser generation manifest itself in different ways. In most of the works the main conclusion about the lasing nature of the emission is made based on the narrow line width and threshold behavior [2-5, 7-16, 18, 19, 21, 22, 25-27, 30, 32-43] of the emission. Rarely the emission polarization was tested [11, 13, 23, 24, 27, 28, 32-37, 39], the direct observation of the light field distribution in the resonator was made [10, 13, 14, 16, 18, 19, 21, 34, 39, 41] or the laser emission directionality was observed [22, 27, 31, 34, 36]. To our best knowledge the intensity correlations measurements above the lasing threshold was demonstrated only in [26], and the spatial coherence of the lasing emission was shown only in [27, 30]. For halide perovskites single crystals both ASE [43] and RL [5] were already demonstrated at temperature 77 K and room temperature respectively. At such temperatures, the exciton resonance experiences a sufficiently large temperature broadening, which complicates the analysis of the spectra. RL in halide perovskites could found its applications in the speckle free imaging [44]. In all the above works either ASE or lasing were demonstrated, and even if both phenomena were reported in the same work, experiments were carried out in different structures.

In this study we present the observation of ASE and RL in the same $MAPbBr_3$ halide perovskite single crystal. A reliable separation of the two emission regimes was achieved by comparing the emission line width, mode structure, the growth of the emission intensity after the threshold, and its polarization degree. Low temperature optical measurements (T = 4 K) allowed us to identify the gain region spectral position at defect-related states below the bound exciton resonance.

## 2. Results and Discussion

MAPbBr$_3$ halide perovskite single crystals were synthesized using the solution method. Figure S1 (a, b) shows the XRD pattern of the single crystal as obtained after the synthesis (a) and of the grounded crystal (b). In the pattern of the single crystal only four reflexes can be observed, which belong to the (001) facet and multiples of that facet. Thus the crystal has a clear orientation without other facets being present. The XRD pattern of the powder is containing only reflexes that can be addressed to methylammonium lead tribromide, assuring the perovskite structure and the phase purity of the material. The orientation of the single crystal can be seen as well in the SEM images (Figure S2). The crystal has a rough surface, and small cracks can be noticed (**Figure 1** (b), marked by arrow). With backside illumination the cracks are visible also in the optical microscope (Figure 1 (a)).

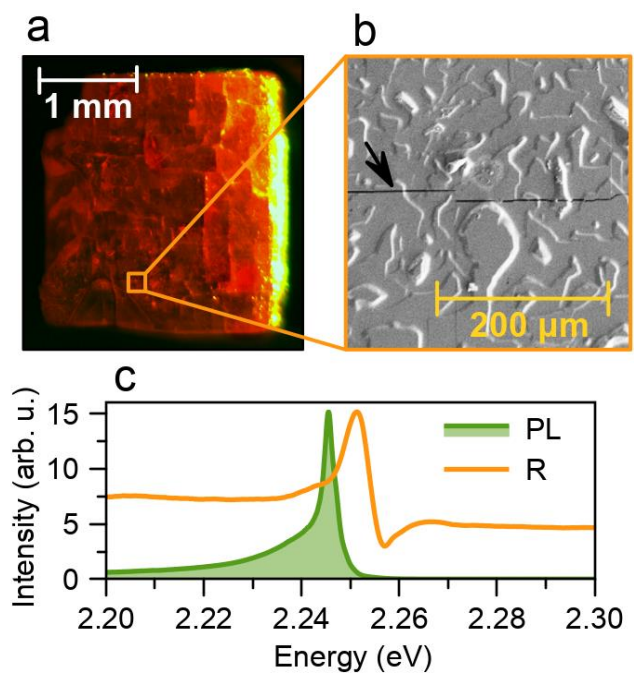

**Figure 1.** Optical microscope image of the MAPbBr$_3$ single crystal (a) and magnified SEM image (b). The black arrow marks a crack. (c) Photoluminescence (PL, green curve) and normal reflectivity (R, orange curve) spectra taken at 4 K.

For the optical emission measurements the MAPbBr$_3$ single crystal was cooled down to a temperature of 4 K. Figure 1 (c) shows PL (green curve) and normal reflectivity (orange curve) spectra. A clear free excitonic transition could be observed in the reflectivity at 2.255 eV. The PL spectra has a narrow bound exciton peak with a Stokes-shift of around 5 meV, and a defect-related broad PL at energies below this peak [45]. For the PL pumping femtosecond UV laser pulses with a 80 MHz repetition rate were used.

With the increase of the pump intensity at the low-energy defect-related tail of the PL spectra a new additional PL signal appears at pump densities per pulse $I_{pump} > 1$ μJ cm$^{-2}$. Depending on the region of the sample, two different types of the additional PL signal can be distinguished. In most regions, a broad PL line with the quality factor Q ~ $10^2$ was observed (**Figure 2** (a) and Figure S3 (left column)). A series of narrow lines with Q~$10^3$ were observed near cracks and crystal boundaries (**Figure 2** (c) and Figure S3 (right column)). We assume that the Figure 2 (a) demonstrates the amplified spontaneous emission (ASE), and the Figure 2 (c) demonstrates the random lasing (RL) phenomenon. The quality factor was determined as Q = $\lambda/\Delta\lambda$, where $\Delta\lambda$ is the line width and $\lambda$ the spectral position of the emission peak.

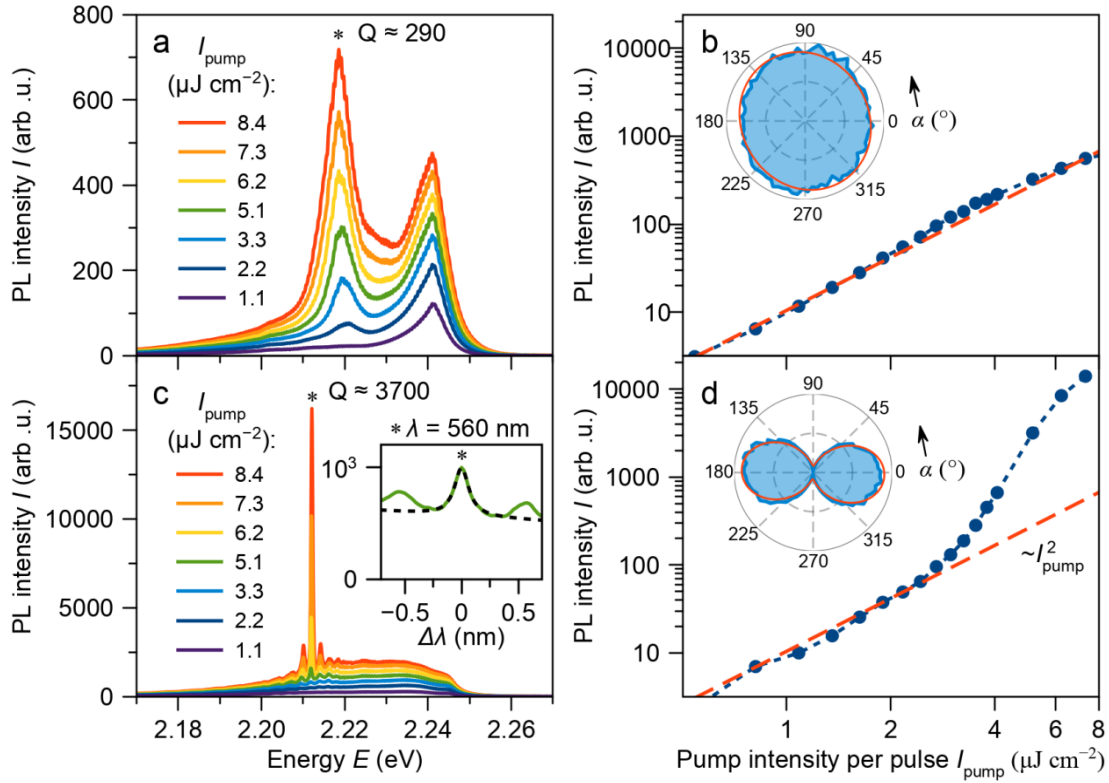

**Figure 2.** Typical PL spectra for different pump densities per pulse $I_{pump}$ for ASE (a) and RL (c). The inset in (c) shows the magnified PL spectrum for $I_{pump}$ = 5.1 μJ cm$^{-2}$ (green curve) and its Lorentzian fit (black dashed curve) ($\Delta\lambda$ – detuning from the peak central wavelength $\lambda$). Log-log dependence of the PL intensity on $I_{pump}$ for ASE (b) and RL (d) for peaks denoted by (*). Dots – experimental data, dashed lines – quadratic fits below the threshold. Spectra are normalized by the PL intensity growth below the threshold. Insets in (b) and (d) show PL polarization diagrams above the threshold for peaks denoted by (*). Blue curves – experimental data, red curves – fits by (2).

ASE process is the spontaneous emission which was optically amplified by the stimulated emission in the perovskite gain medium after the pump pulse. The presence of a threshold is inherent in this process, and a more rapid increase in the PL intensity is expected after the threshold. A clear threshold could be observed in Figure 2 (a) at $I_{pump}$ ~ 2 μJ cm$^{-2}$. In Figure 2 (b), the dots show the PL intensity of the ASE line (marked by (*) in Figure 2(a)), depending on $I_{pump}$. The spontaneous emission intensity growth before the threshold is proportional to the squared pump intensity (dashed line in Figure 2(b)), which is a hallmark of exciton-related luminescence. After the threshold it is succeeded by the slightly faster growth corresponding to the additional stimulated emission. A broad ASE spectrum is determined by the interplay of the gain and absorption spectra of the media. Surprisingly this emission is well below the bound exciton resonance indicating that the main role is played by the stimulated emission from defect-related states in the transparency region of the crystal.

An essentially different behavior is observed in RL regions, where the random resonator is formed by cracks and inhomogeneities of the sample crystal. Due to the multiple passage of emitted light in the resonator with the gain medium, the emission spectrum above the threshold is dominated by a set of narrow modes (Figure 2 (c) and Figure S4). The spectral envelope of these modes is still close to the ASE spectrum, but the Q-factors of individual modes are 1-2 orders of magnitude greater than in the ASE regime. The inset in the Figure 2 (c) shows the dominant lasing mode and its Lorentzian fit yielding to the Q ~ 3700.

Another feature of the RL modes is the much faster intensity growth after the threshold, compared to the ASE. Figures 2 (b) and (d) are normalized to the same emission growth with the same quadratic fit shown by the red dashed line. The S-shaped curve could be seen for the RL case, which is typical for lasing.

A direct comparison of the emission growth after the threshold is not entirely correct, since the result will also depend on the directional diagram of the emission. However, in the ASE regime, it is expected that the emission is non-directional. Therefore, faster emission intensity growth in the RL regime indicates a greater gain due to multiple light passages of the resonator.

The emission polarization allows to distinguish ASE and RL regimes. The inset in the Figure 2 (b) shows the polarization diagram for the ASE emission at the spectral position marked by (*) with the non-polarized spontaneous emission background extracted. This stimulated emission is almost non-polarized which proves the absence of the dominant optical mode in ASE. In the RL regime the same diagram (Figure 2 (d)) shows that this emission is polarized. The polarization degree could be estimated as:

$$P = \frac{I_{max} - I_{min}}{I_{max} + I_{min}} \ (1),$$

where $I_{max}$ and $I_{min}$ are the maximum and minimum nonlinear PL intensities at different polarizations. For data presented in Figure 2 in the case of ASE this parameter is $P \sim 0.06$ indicating the non-polarized emission, for RL it is $P \sim 0.96$ indicating polarized lasing. The red dashed curves in Figure 2 (b,d) insets show fittings of the polarization diagrams by the equation:

$$I(\alpha) = I_0 \left( \cos^2 \alpha + \frac{1-P}{1+P} \sin^2 \alpha \right) (2),$$

where is the analyzer rotation angle, representing the combination of the non-polarized background and the polarized emission determined by the Malus's law.

Table S4 summarizes emission parameters for different sample points with the ASE and RL regimes. It can be seen that the described ASE and RL properties are sufficiently universal and could be used to clearly distinguish these two regimes.

## 3. Conclusions

In this study we have shown that emission from the MAPbBr$_3$ single crystal induced by pulsed UV pumping at 4 K can be divided in two well distinct groups: ASE and RL, depending on the emission intensity growth after the threshold, quality factor and emission polarization degree. Moreover, utilization of the low temperature (4 K) let us observe the fine structure of the emission. The formation of resonators for RL is a random process, however, it helps to study the nature of laser generation in the highest quality single crystal material. Both types of the emission were found to be spectrally located below the bound exciton transition in the defect-related emission region. Therefore, our observations of emission spectral localization and its fine structure raise new questions about the underlying emission mechanisms, which are important for deeper understanding of the gain medium properties of these novel halide perovskite materials. These questions could be resolved only in further low temperature studies of halide perovskite single crystals by photoluminescence excitation spectroscopy and other methods.

**Experimental Section**

*Chemicals*: Methylammonium bromide (98 %, Sigma Aldrich), lead bromide ($\geq$ 98 %, Sigma Aldrich) and dimethylformamide (DMF) were used as received without further purification.

*Synthesis*: 1M solutions of methylammonium bromide and lead bromide were prepared by dissolving both compounds in 3 ml DMF. After stirring for some minutes the solution was filtrated using a 0.2 μm PTFE filter. The solution was transferred into a vial, which was closed with a glass cap and sealed with Parafilm. Afterwards the vial was placed in a beaker and heated in a water/ethanol bath to 60°C and the temperature was kept constant overnight. On the next day the temperature was increased to 80°C and after two hours of further heating the crystals were obtained (< 15 crystals). The crystals have been dried and used without further treatment for the measurements. The XRD measurements were performed with a Bruker D8 Advance diffractometer with the Cu Kα radiation. Scanning electron microscopy images were obtained using a Zeiss Crossbeam 1540XB workstation.

*Optical characterization*: The MAPbBr$_3$ single crystal was mounted into a closed-cycle helium cryostat Montana Cryostation, and kept at 4 K during the experiment. The sample was excited by the frequency doubled light of a Ti:Sapphire laser (Spectra Physics MaiTai) with 150 fs pulses and a repetition rate of 80 MHz. The excitation wavelength was 370 nm. The laser beam was focused on the sample into a 8 μm spot by a 20x Mitutoyo micro objective lens. The same lens was used to collect the PL. PL spectra were measured using the spectrometer with a grating 1200 mm$^{-1}$ equipped with the linear CCD-detector. An achromatic λ/2 plate and a Glan-Taylor prism were mounted in front of the spectrometer slit in order to measure the PL polarization.

**Supporting Information**
Supporting Information is available from the Wiley Online Library or from the author.

**Acknowledgements**
This study was supported by the Russian Science Foundation (project 19-72-10034). C.G. thank the Gottfried Wilhelm Leibniz University Hannover for financial support (Wege in die Forschung II). The authors thank Dr. Mariano Curti for performing the XRD measurements. This work was carried out on the equipment of SPbU Resource center "Nanophotonics" and partly using research facilities of laboratory "Photoactive nanocomposite materials" supported within SPbU program (ID: 51124539).

Received:      ((will      be      filled      in      by      the      editorial      staff))
Revised:      ((will      be      filled      in      by      the      editorial      staff))
Published online: ((will be filled in by the editorial staff))

TOC entry:

Amplified spontaneous emission and random lasing regimes are differentiated in the photoluminescence from MAPbBr₃ single crystals at 4 K and pulsed UV pumping. These two regimes are differentiated by the line width, mode structure and polarization of the emission, and its growth after the threshold.

Aleksei O. Murzin[*], Boris V. Stroganov, Carsten Günnemann, Samia Ben Hammouda, Anna V. Shurukhina, Maxim S. Lozhkin, Alexei V. Emeline, Yury V. Kapitonov[**]

**Title** Amplified Spontaneous Emission and Random Lasing in MAPbBr₃ Halide Perovskite Single Crystals

ToC figure

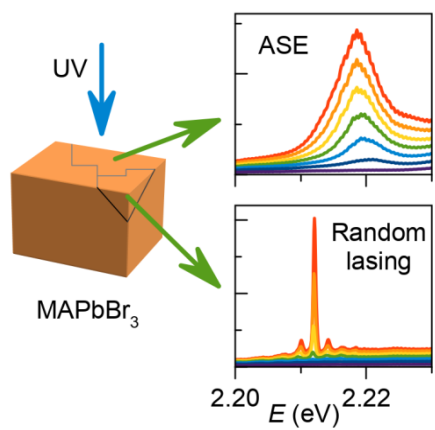

## Supporting Information

**Title** Amplified Spontaneous Emission and Random Lasing in MAPbBr$_3$ Halide Perovskite Single Crystals

*Aleksei O. Murzin[*], Boris V. Stroganov, Carsten Günnemann, Samia Ben Hammouda, Anna V. Shurukhina, Maxim S. Lozhkin, Alexei V. Emeline, Yury V. Kapitonov[**]*

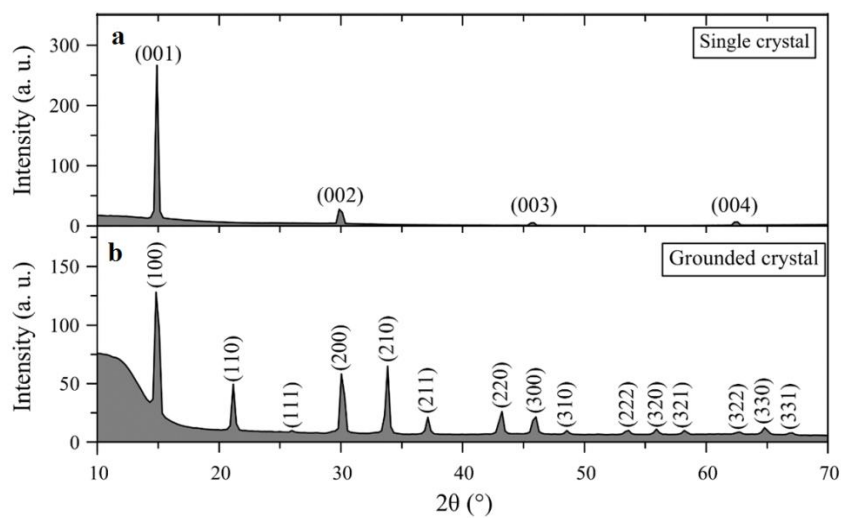

**Figure S1.** XRD patterns of the single (a) and grounded (b) MAPbBr$_3$ crystals.

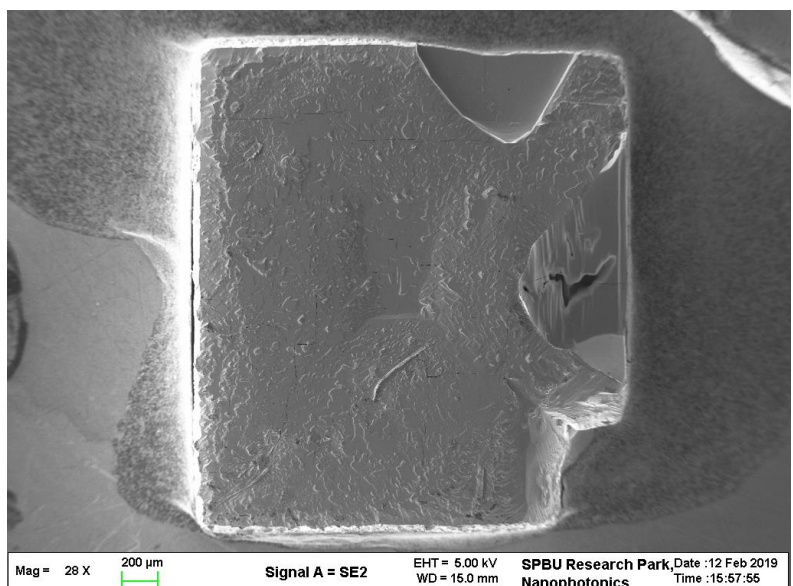

**Figure S2.** SEM image of the MAPbBr$_3$ single crystal.

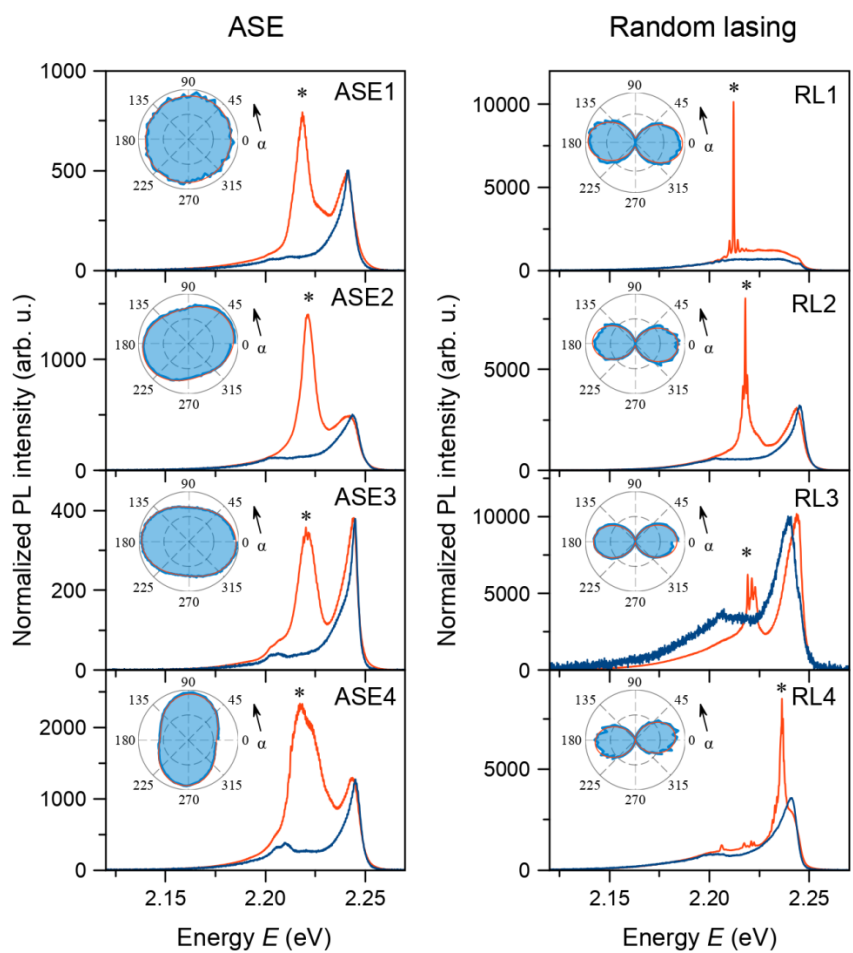

**Figure S3.** More examples of PL spectra with the ASE (left column) and RL (right column) regimes from different sample points. For each panel the blue (red) curve corresponds to the spectrum below (above) the threshold. Spectra are normalized to the highest energy emission peak intensity. Insets show PL polarization diagrams above the threshold for peaks denoted by (*). Blue curves – experimental data, red curves – fits by (2) from the main text.

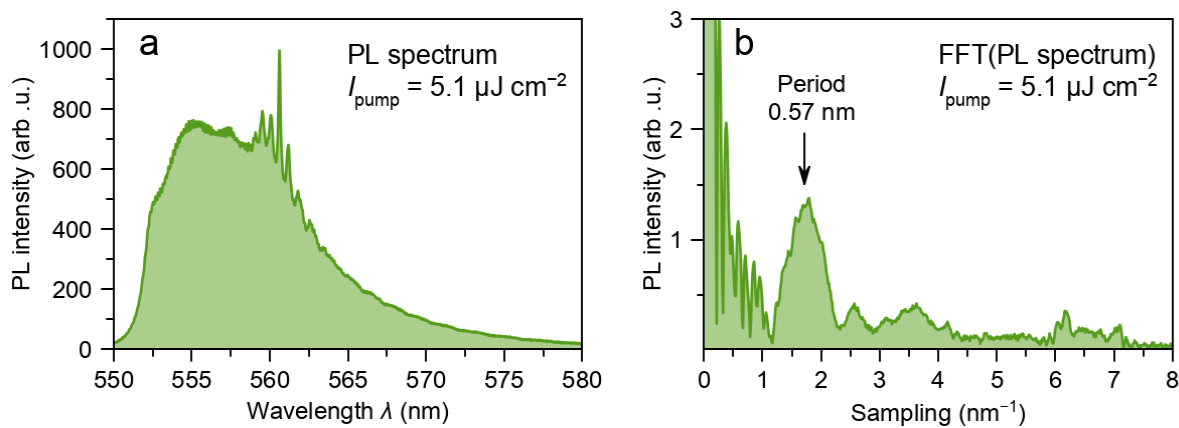

**Figure S4.** Emission spectra (a) and its FFT (b) for $I_{pump}$=5.1 µJ cm⁻² for the sample point from Figure 2(c). Arrow shows the periodical component corresponding to the RL modes.

| Sample point | Threshold, $\mu J\ cm^{-2}$ | Q-factor | Polarization degree P | Intermodal distance, meV |
|---|---|---|---|---|
| ASE1 | 1.9 | 291 | 0.02 | – |
| ASE2 | 2.2 | 359 | 0.14 | – |
| ASE3 | 2.3 | 226 | 0.16 | – |
| ASE4 | 1.4 | 252 | 0.24 | – |
| RL1 | 4.5 | 3718 | 0.83 | 2.2 |
| RL2 | 3.7 | 4752 | 0.92 | 1.1 |
| RL3 | 3.5 | 2725 | 0.96 | 1.0 |
| RL4 | 0.6 | 5371 | 1.00 | 0.5 |

**Table S4.** Nonlinear PL parameters for peaks denoted by (*) in Figure S3.